\documentclass[11pt]{amsart} 
\usepackage{amscd,amssymb,amsxtra}
\usepackage{mathrsfs}  % \mathrsfs{A}
\DeclareSymbolFontAlphabet{\mathrsfs}{rsfs}
\usepackage[mathscr]{eucal} % \mathscr{A} 
\usepackage{comment}
\usepackage{color}
\usepackage{enumitem} 
\usepackage[colorlinks,backref=page,citecolor=refkey]{hyperref}
\usepackage{aliascnt}

\usepackage{marginnote}

\makeatletter
\let\@secnumfont\bfseries
\def\section{\@startsection{section}{1}%
  \z@{4\linespacing\@plus\linespacing}{\linespacing}%
  {\bfseries\centering}}
\def\introsection{\@startsection{section}{1}%
  \z@{3\linespacing\@plus\linespacing}{\linespacing}%
  {\bfseries\centering}}
\def\subsection{\@startsection{subsection}{2}%
   \z@{1.25\linespacing\@plus.7\linespacing}{.5\linespacing}%
   {\normalfont\bfseries}}
\def\subsectionsinline{\def\subsection{\@startsection{subsection}{2}%
  \z@{1\linespacing\@plus.7\linespacing}{-.5em}%
  {\normalfont\bfseries}}}

\makeatother

\numberwithin{equation}{section}

\newcommand{\mynewtheorem}[2]{
  \newaliascnt{#1}{equation}
  \newtheorem{#1}[#1]{#2}
  \aliascntresetthe{#1}
  \expandafter\def\csname #1autorefname\endcsname{#2}
}

\theoremstyle{definition}
\mynewtheorem{definition}{Definition}
\mynewtheorem{example}{Example}
\mynewtheorem{problem}{\color{blue}Problem}
\mynewtheorem{probsec}{\color{blue}Problem}
\mynewtheorem{exercise}{Exercise}
\mynewtheorem{question}{\color{blue}Question}
\mynewtheorem{project}{\color{blue}Project}
\mynewtheorem{construction}{Construction}
\mynewtheorem{task}{\color{blue}Task}
\mynewtheorem{otask}{\color{green}Obsolete Task}
\mynewtheorem{notation}{Notation}

\newtheorem*{definition*}{Definition}
\newtheorem*{example*}{Example}
\newtheorem*{problem*}{\color{blue}Problem}
\newtheorem*{probsec*}{\color{blue}Problem}
\newtheorem*{exercise*}{Exercise}
\newtheorem*{question*}{\color{blue}Question}
\newtheorem*{project*}{\color{blue}Project}
\newtheorem*{construction*}{Construction}
\newtheorem*{notation*}{Notation}

\theoremstyle{remark}
\mynewtheorem{note}{Note}
\mynewtheorem{remark}{Remark}
\mynewtheorem{data}{Data}

\newtheorem*{note*}{Note}
\newtheorem*{remark*}{Remark}
\newtheorem*{data*}{Data}

\theoremstyle{plain}
\mynewtheorem{theorem}{Theorem}
\mynewtheorem{corollary}{Corollary}
\mynewtheorem{lemma}{Lemma}
\mynewtheorem{proposition}{Proposition}
\mynewtheorem{conjecture}{Conjecture}
\mynewtheorem{claim}{Claim}
\mynewtheorem{proposal}{Proposal}
\mynewtheorem{conclusion}{Conclusion}
\mynewtheorem{hypothesis}{Hypothesis}
\mynewtheorem{assumption}{Assumption}

\newtheorem*{theorem*}{Theorem}
\newtheorem*{corollary*}{Corollary}
\newtheorem*{lemma*}{Lemma}
\newtheorem*{proposition*}{Proposition}
\newtheorem*{conjecture*}{Conjecture}
\newtheorem*{claim*}{Claim}
\newtheorem*{proposal*}{Proposal}
\newtheorem*{conclusion*}{Conclusion}
\newtheorem*{hypothesis*}{Hypothesis}
\newtheorem*{assumption*}{Assumption}

\newenvironment{proof*}[1][\proofname]{
  \begin{proof}[#1]}{  
\end{proof}}

\definecolor{refkey}{rgb}{0,.6,.4}

\newcommand\mybigwedge{\raisebox{2pt}{\scalebox{.9}{$\bigwedge$}}}
\renewcommand{\:}{\colon}
\newcommand{\CC}{{\mathbb C}}

\DeclareMathOperator{\End}{End}

\DeclareMathOperator{\pt}{pt}

\newcommand{\RR}{{\mathbb R}}

\newcommand{\ZZ}{{\mathbb Z}}

\newcommand{\chiup}{\raise.5ex\hbox{$\chi$}}
\newcommand{\cir}{S^1}

\DeclareRobustCommand{\mstrut}{^{\vphantom{1*\prime y\vee M}}}

\newcommand{\temsquare}{\raise3.5pt\hbox{\boxed{ }}}

\newcommand{\zmod}[1]{\ZZ/#1\ZZ}

\newcommand{\zt}{\zmod2}

\newcommand{\hneg}{\mkern-.5\thinmuskip}
 
\DeclareFontFamily{U}{mathx}{}
\DeclareFontShape{U}{mathx}{m}{n}{<-> mathx10}{}
\DeclareSymbolFont{mathx}{U}{mathx}{m}{n}
\DeclareMathAccent{\widehat}{0}{mathx}{"70}
\DeclareMathAccent{\widecheck}{0}{mathx}{"71}
 
\DeclareMathSymbol{\bigtimes}{1}{mathx}{"91}

\usepackage[all,2cell]{xy}\renewcommand{\cir}{\ensuremath{S^1}}
\usepackage{graphicx}\usepackage{epsf}
\definecolor{refkey}{rgb}{0,.8,.2}\definecolor{labelkey}{rgb}{1,0,0} 

\DeclareMathOperator{\Arf}{Arf}
\DeclareMathOperator{\Cliff}{Cliff}
\DeclareMathOperator{\Pfaff}{Pfaff}
\DeclareMathOperator{\Sym}{Sym}
\DeclareMathOperator{\pfaff}{pfaff}
\DeclareMathOperator{\trace}{trace}
\newcommand{\CWCs}{\Cliff(\WCs)}
\newcommand{\Clc}{\textnormal{C}\ell^{\CC}}

\newcommand{\DX}{D\mstrut _{\hneg X}}
\newcommand{\WCs}{W^*_{\CC}}
\newcommand{\mX}{\mu \mstrut _{\hneg X}}
\newcommand{\tX}{\widetilde{X}}
\newcommand{\tps}{\tilde\psi }

\begin{document}

\abovedisplayskip18pt plus4.5pt minus9pt
\belowdisplayskip \abovedisplayskip
\abovedisplayshortskip0pt plus4.5pt
\belowdisplayshortskip10.5pt plus4.5pt minus6pt
\baselineskip=15 truept
\marginparwidth=55pt

\makeatletter
\renewcommand{\tocsection}[3]{%
  \indentlabel{\@ifempty{#2}{\hskip1.5em}{\ignorespaces#1 #2.\;\;}}#3}
\renewcommand{\tocsubsection}[3]{%
  \indentlabel{\@ifempty{#2}{\hskip 2.5em}{\hskip 2.5em\ignorespaces#1%
    #2.\;\;}}#3} 
\renewcommand{\tocsubsubsection}[3]{%
  \indentlabel{\@ifempty{#2}{\hskip 5.5em}{\hskip 5.5em\ignorespaces#1%
    #2.\;\;}}#3} 
%  \indentlabel{\hskip 4em#3}}
\def\@makefnmark{%
  \leavevmode
  \raise.9ex\hbox{\fontsize\sf@size\z@\normalfont\tiny\@thefnmark}} 
\def\multfoot{\textsuperscript{\tiny\color{red},}}
\def\footref#1{$\textsuperscript{\tiny\ref{#1}}$}
%             \footnote{Note\label{SYM}} ... in footnote~\footref{SYM}
\makeatother

\renewcommand{\theremark}{\arabic{remark}}
\renewcommand{\theexample}{\arabic{example}}
\renewcommand{\theequation}{\arabic{equation}} 

\setcounter{tocdepth}{2}

%**end of header

% lasteq@  6
% lastsec@  1
% lastthm@  6
% lastfig@  4

 \title[The Odd Fermion]{The Odd Fermion} 

 \author[D. S. Freed]{Daniel S.~Freed}
 \address{Harvard University \\ Department of Mathematics \\ Science Center
Room 325 \\ 1 Oxford Street \\ Cambridge, MA 02138}
 \email{dafr@math.harvard.edu}
 
 \author[M. J. Hopkins]{Michael J.~Hopkins}
 \address{Harvard University \\ Department of Mathematics \\ Science Center
Room 325 \\ 1 Oxford Street \\ Cambridge, MA 02138}
 \email{mjh@math.harvard.edu}

 \author[C. Teleman]{Constantin Teleman} 
  \address{Department of Mathematics \\ University of California \\ 970 Evans
 Hall \#3840 \\ Berkeley, CA 94720-3840}  
  \email{teleman@berkeley.edu}

 \thanks{This material is based upon work supported by the National Science
Foundation under Grant Number DMS-2005286 and by the Simons Foundation Award
888988 as part of the Simons Collaboration on Global Categorical Symmetries.
This work was performed in part at Aspen Center for Physics, which is supported
by National Science Foundation grant PHY-2210452.}
% \dedicatory{}
 \date{November 28, 2023}
 \begin{abstract} 
 In this short note we use the geometric approach to (topological) field theory
to address the question: Does an odd number of quantum mechanical fermions make
sense? 
 \end{abstract}
\maketitle

\emph{Old questions}: Does an odd number of quantum mechanical fermions make
sense?  And, if so, what is the dimension of the Hilbert space of states?

\emph{Answers}: Yes, it makes sense as an anomalous theory.  The standard
definition of `dimension' gives~0, and the ``twisted dimension'' for a single
fermion is~$\sqrt2$.
 
Key points:

  \begin{itemize}

 \item The theory is anomalous: it is a 1-dimensional field theory relative to
the 2-dimensional \emph{Arf theory}.

 \item The theory is topological, so computations are easily carried out.

 \item The state space is a module over the Clifford algebra~$\Clc_1$;
`dimension' and `twisted dimension' are relative concepts---relative to the
Clifford algebra action.

 \item The tensor product construction of a theory of $N_1+N_2$ fermions works
as it should.

  \end{itemize}

\noindent
 This question has garnered some interest over the years, and the theory of an
odd number of fermions has sometimes been deemed inconsistent.  See \cite{DGG},
\cite[\S2.1]{W}, and \cite[footnote~1]{SS} for recent comments; the latter
contains several additional references to the literature.  That the state space
is to be regarded as a module over~$\Clc_1$ already appears
in~\cite[Remark~5.10]{F1}.
 
We thank Nati Seiberg for his perspective.  His lecture at the November, 2023
annual meeting of the Simons Collaboration on Global Categorical Symmetries
inspired us to write this note.

%   \section{}\label{sec:1}
% lastsubsec@  2

  \subsection*{Quantum mechanics of $N$ fermions}\label{subsec:1.1}

Let $N$~be a positive integer, which we presently assume is odd.  Consider the
quantum mechanical theory of $N$~free massless spinor fields.  It is a
1-dimensional theory which, after Wick rotation, has background fields a
Riemannian metric and spin structure.  The fluctuating field is a spinor field
with values in a real vector space of dimension~$N$.  For $N$~even, one can
work out this theory as a standard exercise; see \cite[FP16]{F2} for example.

  \begin{remark}[]\label{thm:1}
 We assume from the beginning that the spinor field is fermionic, or odd
(anticommuting).  Therefore, its quantization generates fermionic states.
Furthermore, Bose-Fermi statistics are encoded in a $\zt$-grading on the state
space, and algebras of observables are also $\zt$-graded; they are
\emph{superalgebras}.  Alternatively, one can consider the system in which the
spinor field is bosonic and there are no fermionic states; see the discussion
of a lattice version in~\cite[\S1.1]{SS}.
  \end{remark}

Let $W$~be an $N$-dimensional real inner product space.  The spin structure on
a spin Riemannian 1-manifold~$X$ is given as a double cover $\tX\to X$.  A
spinor field~$\psi $ on~$X$ is a function $\tps\:\tX\to \Pi W$ that is odd
under the deck transformation of $\tX\to X$.  Here $\Pi W$~is the
parity-reversed vector space, i.e., $W$~as an odd vector space.\footnote{More
precisely, $\Pi W=\Pi \otimes W$, where $\Pi $~is the canonical odd line.}  The
lagrangian density is
  \begin{equation}\label{eq:1}
     L = \frac 12\langle \psi ,\DX \psi \rangle\,d\mX, 
  \end{equation}
where $\mX$~is the Riemannian measure on~$X$ and $\DX$~is the Dirac operator
on~$X$.  The latter is $\DX\psi =\partial \tps$ for $\partial $~the positively
oriented unit norm vector field on~$\tX$.
 
Compute canonical quantization on~$Y=\pt$ as follows.  First, solve the
classical equations of motion on~$\RR\times Y$ to obtain a symplectic vector
space, which in this case is odd.  Here the classical equation asserts that
$\psi $~is locally constant, so the vector space of solutions is canonically
isomorphic to~$\Pi W$.  After complexification, classical observables form the
algebra $\Sym^{\bullet }\Pi \WCs$.  This is the finite dimensional exterior
algebra of~$\WCs$ with the $\zt$-grading that makes $\mybigwedge ^1\WCs=\WCs$
odd.  Second, form the \emph{deformation quantization} using the Poisson
bracket derived from the lagrangian~\eqref{eq:1}.  The result is the Clifford
algebra $\Cliff(\WCs)$ for the complexified dual inner product; the Clifford
algebra carries its usual $\zt$-grading.  Third, quantum states are defined by
producing an irreducible $\zt$-graded module~$M$ over $\CWCs$.  Here is where
the parity of~$N$ shows up.  If $N$~is even, then the complex algebra $\CWCs$
of observables is the full endomorphism algebra of~$M$.  In this case $M$~is a
$\zt$-graded vector space of dimension
  \begin{equation}\label{eq:2}
     \dim M = 2^{N/2-1}\bigm| 2^{N/2-1}; 
  \end{equation}
the notation `$b|f$' indicates the dimensions of the even (Bose) and odd
(Fermi) homogeneous subspaces of a $\zt$-graded vector space.  For $N$~odd the
action of~$\CWCs$ on~$M$ has a nontrivial \emph{commutant}:\footnote{As always
in super algebra, the commutant is taken in the $\zt$-graded sense: odd
elements $a_1,a_2$ commute if $a_2a_1=-a_1a_2$.} the subalgebra of~$\End(M)$
consisting of operators that commute with every element of $\CWCs$ is
isomorphic to $\Clc_1$, the complex Clifford algebra on a single odd
generator~$\delta $. 

  \begin{example}[$N=1$]\label{thm:2}
 Let $W=\RR$ with its standard inner product, so $\CWCs=\Clc_1$ is the complex
superalgebra generated by an odd element~$\gamma $.  An irreducible module is
$M=\CC^{1|1}$ with $\gamma =\left(\begin{smallmatrix}
0&1\\1&0  \end{smallmatrix}\right)$.  The commutant is generated by $\delta
=\left(\begin{smallmatrix} 0&-i\\i&0  \end{smallmatrix}\right)$.  
  \end{example}

Key point: One must consider the state space as a module over the commutant.
Explanation: In this case the algebra of complex observables~$A$ is the
commutant of the commutant~$B$ inside the full endomorphism algebra~$\End(M)$,
so is identified as $\End_B(M)$, the algebra of endomorphisms of~$M$ as a
$B$-module.  Typically, one must complete the module~$M$ to obtain a Hilbert
space of states, but in this case $M$~is finite dimensional and no completion
is necessary.

  \begin{remark}[]\label{thm:6}
 An extra ``spectator fermion'' has appeared in many previous works, for
example~\cite{tH,LW,K}.  It plays a similar role to the commutant, but the
construction in the previous paragraph applies generally and leads to the
commutant as part of the anomaly.
  \end{remark}

Next, we compute (heuristically, using a regularization) the path integral
on~$X=\cir$.  For simplicity take~$N=1$, as in \autoref{thm:2}; the general
case is the $N^{\textnormal{th}}$~power of the computation here.  There are two
spin structures: \emph{bounding} (antiperiodic, Neveu-Schwarz) and
\emph{non-bounding} (periodic, Ramond).  The relevant one for us is the
bounding spin structure.  Let $X$~be a Riemannian circle of length~$L$.  The
path integral of the Lagrangian~\eqref{eq:1} is the pfaffian~$\pfaff \DX$ of
the Dirac operator.  It is naturally an element of the real Pfaffian
line~$\Pfaff \DX$.  That line carries a natural inner product, and the norm
square of $\pfaff\DX$ is the determinant of~$\DX$.  The set of eigenvalues
of~$\DX$ is
  \begin{equation}\label{eq:3}
     \left\{ \frac{2\pi in}{L}  : n\in \ZZ+\frac 12\right\}.
  \end{equation}
The computation of the product of eigenvalues is 
  \begin{equation}\label{eq:4}
     \prod\limits_{n\in\ZZ+1/2}\left(\frac{2\pi i}{L}\,n\right)
     =\prod\limits_{n\in\ZZ+1/2}n=\prod\limits_{\text{$n$
     odd}}n=\frac{\prod\limits_{n\not=0}n}{\prod\limits _{ \substack{\text{$n$
     even}\\ n\not=0}}  n}=\frac{\prod\limits_{n\not=0}n}
     {\prod\limits_{n\not=0}2n}=2. 
  \end{equation}
We use $\zeta $-functions to regulate the infinite products.  In the first and
second steps we use that the $\zeta $-regularized cardinality of~$\ZZ$, and so
too of~$\ZZ+1/2$, is zero.  In the last step we use the corollary that the
$\zeta $-regularized cardinality of $\ZZ^{\neq 0}$ is~$-1$.  The pfaffian is
the square root of~\eqref{eq:4}, determined here up to sign.

  \begin{remark}[]\label{thm:3}
 \ 
 \begin{enumerate}[label=\textnormal{(\arabic*)}]

 \item The path integral is independent of the length~$L$: the theory is
\emph{topological}. 

 \item That the state space~$M$ is a module over~$\Clc_1$ signals that the
theory is \emph{anomalous}, as confirmed by the fact that the path integral is
an element of a line rather than a number. 
 
 \item In the next section we use a canonical element of~$\Pfaff\DX$ to
identify the pfaffian $\pfaff\DX$ as a number: $\sqrt2$.

 \item For the nonbounding spin structure the Dirac operator has a
1-dimensional kernel, and hence the $\zt$-graded Pfaffian line is odd in the
sense of super vector spaces.  The nonzero kernel also implies that the
pfaffian element is zero.  In fact, this is a consequence of the oddness of the
Pfaffian line: in the category of super vector spaces every homomorphism
$\CC\to \Pi $ from an even line to an odd line is zero.

 \end{enumerate}
  \end{remark}

  \subsection*{$N$ fermions as a boundary topological field theory}\label{subsec:1.2}

In the Wick-rotated setting an anomaly is a once-categorified invertible field
theory, and the anomalous theory is defined relative to it.  Both theories are
defined on the same bordism category, so in particular have the same
dimension.  It often happens that the anomaly extends to a full theory in one
higher dimension.  See~\cite{F3} for discussion and motivation.   

  \begin{figure}[ht]
  \centering
  \includegraphics[scale=1]{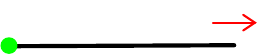}
  \vskip -.5pc
  \caption{The $\Clc_1$-module $M$}\label{fig:1}
  \end{figure}

Fix $N$~to be odd.  We work with topological field theories defined on spin
manifolds and taking values in the Morita 2-category of complex superalgebras.
The preceding implies that the anomaly theory of $N$~fermions assigns the
algebra\footnote{The Clifford algebra is invertible in the Morita 2-category of
superalgebras.  This is equivalent to its being \emph{central simple}.
See~\cite{Wa,De}.}~$\Clc_1$ to a point and the Pfaffian line of Dirac to a
circle.  The resulting once-categorified 1-dimensional invertible
theory~$\alpha $ is the truncation of the 2-dimensional invertible ``Arf
theory''; see~\cite{Gu,DeGu} for example.  The partition function on a closed
spin 2-manifold~$W$ is $\alpha (W)=(-1)^{\Arf W}$, where $\Arf W$ is the Arf
invariant of the spin structure: 0~for even spin structures and 1~for odd spin
structures.  The quantum theory of $N$~fermions is a boundary theory of~$\alpha
$; the combined theory ($\alpha \;+$ boundary) assigns the left
$\Clc_1$-module~$M$ to the bordism depicted in \autoref{fig:1}.  Green depicts
components of the boundary colored by the module theory, and the red arrow
tells if an uncolored boundary component is incoming or outgoing.
 
In a spin theory the usual dimension of the value of the theory on a
codimension one closed manifold~$Y$ is computed by taking Cartesian product
of~$Y$ with the \emph{nonbounding}~$\cir$: the spin double cover of the
nonbounding~$\cir$ is the product double cover, so it computes the trace of the
identity map in whatever category we are in.  Furthermore, this trace lives in
the Hochschild \emph{homology} of the value of~$Y$.  In our case, for $Y=\pt$
the Arf theory on the Cartesian product evaluates to $\alpha
(\cir_{\textnormal{nonbounding}})=\Pi $, where $\Pi $~is the odd line.  As
mentioned in \autoref{thm:3}(4), an odd line has no nonzero elements.  Hence
the usual dimension is~0.

  \begin{remark}[]\label{thm:4}
 This is the `usual dimension', or categorical dimension, in the category of
$\Clc_1$-modules.  By comparison, the `usual dimension' in the category of
super vector spaces is $\dim \CC^{b|f}=b-f$.  This is oft referred to as
`$\trace\,(-1)^F$'.  The `twisted dimension' we are after is the categorical
trace of the grading operator, which for the super vector space~$\CC^{b|f}$
equals $b+f$.  We now compute `twisted dimension' in the category of
$\Clc_1$-modules.
  \end{remark}

  \begin{figure}[ht]
  \centering
  \includegraphics[scale=1]{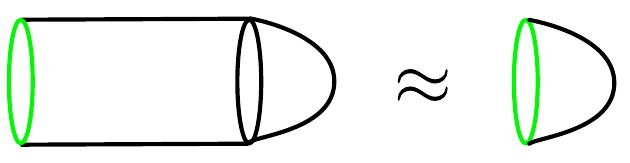}
  \vskip -.5pc
  \caption{The twisted dimension of~$M$}\label{fig:2}
  \end{figure}

Consider the Cartesian product of the bordism in \autoref{fig:1} with the
\emph{bounding} circle~$\cir_{\textnormal{bounding}}$.  The theory ($\alpha
\;+$ boundary) computes the supertrace of the grading operator on~$M$---what we
are calling the \emph{twisted dimension} of~$M$---which lives in the Hochschild
\emph{cohomology}~$\alpha (\cir_{\textnormal{bounding}})$ of the
algebra~$A=\Clc_1$.  The Hochschild cohomology is an algebra, so has a
canonical element: the unit for multiplication.  Since the Arf theory~$\alpha $
is invertible, the Hochschild cohomology is a line, and so the unit is a basis
element.  We use this basis element to define the twisted dimension as a
number, the value of the theory on \autoref{fig:2}.  Compute the square of this
number by evaluating the squared theory~$\alpha ^{\otimes 2}$ on
\autoref{fig:2}.  Observe that $\alpha ^{\otimes 2}(\pt)=\Clc_1\otimes
\Clc_1\cong \Clc_2$, and the square of ($\alpha \;+$ boundary) assigns the left
$\Clc_2$-module $M\otimes M$ to the bordism in \autoref{fig:1}.  Observe that
every finite dimensional $A=\Clc_1$-module is free, so has the form $M=A\otimes
V$ for an even vector space~$V$.  For $N$~fermions ($N$~odd) we have $\dim
V=2^{(N-1)/2}$.  The Morita equivalence $\Clc_2\simeq \CC$ identifies
$\Clc_2$-modules with super vector spaces; under this equivalence $M\otimes M$
corresponds to the super vector space $V\otimes V\otimes A$.  By
\autoref{thm:4}, the twisted dimension of $M\otimes M$ is $2(\dim V)^2$.  This
is the square of the twisted dimension of~$M$, hence the twisted dimension
of~$M$ equals
  \begin{equation}\label{eq:6}
     \sqrt2\,\dim V= 2^{(N-1)/2}\sqrt2. 
  \end{equation}
For a single fermion~$N=1$, this twisted dimension equals~$\sqrt2$.

  \begin{remark}[]\label{thm:5}
 \ 
 \begin{enumerate}[label=\textnormal{(\arabic*)}]

 \item Tensor products work as they should.  For example, the tensor product of
the state spaces in two theories with $N$~odd is the tensor product of two
$\Clc_1$-modules, which is a $\Clc_2$-module.  As above, apply the Morita
equivalence $\Clc_2\simeq \CC$ to identify $\Clc_2$-modules with super vector
spaces.  The twisted dimension of the resulting super vector space is $1/2$~the
product of the twisted dimensions of the $\Clc_1$-modules.

 \item There are variants of this story on pin manifolds.  In that case the
anomaly theory is more involved; see~\cite{DeGu}.

 \end{enumerate}
  \end{remark}

%\appendix

 \bigskip\bigskip
\providecommand{\bysame}{\leavevmode\hbox to3em{\hrulefill}\thinspace}
\providecommand{\MR}{\relax\ifhmode\unskip\space\fi MR }
\providecommand{\MRhref}[2]{%
  \href{http://www.ams.org/mathscinet-getitem?mr=#1}{#2}
}
\providecommand{\href}[2]{#2}

  \end{document}